**Emotional Multifaceted Feedback on AI Tool Use in EFL Learning Initiation: Chain-Mediated Effects of Motivation and Metacognitive Strategies in an Optimized TAM Model**

Le Yao[1] & Yantong Liu[2],*

1. Le Yao; Department of English Language and Literature, Sookmyung Women's university, Seoul, 04510, Republic of Korea{2132271@sookmyung.ac.kr};

2. Yantong Liu; Department of Computer and Information Engineering, Kunsan National University, Gunsan, 54150, Republic of Korea {lyt1994@kunsan.ac.kr}; https://orcid.org/0009-0003-5880-0190

Corresponding:Yantong Liu, lyt1994@kunsan.ac.kr

**Abstract**

This study specifically investigates the initiation phase of EFL learners' engagement with AI tools, focusing on how technology acceptance constructs—perceived usefulness (PU), perceived ease of use (PEOU), and perceived self-efficacy (PSE)—influence learning resilience. Drawing on an optimized Technology Acceptance Model (TAM) and integrating constructs from positive psychology, the study examines the chain-mediated effects of learning motivation (LM) and metacognitive strategies (MS) on resilience outcomes, operationalized through optimism (OP), psychological resilience (PR), and growth mindset (GM). A survey of first-year English majors (N = 730) was conducted, and structural equation modeling was employed to analyze the data. The findings indicate that favorable perceptions of AI tools are significantly associated with enhanced LM and MS, which in turn positively impact resilience measures. These results suggest that the interplay between technology acceptance and internal regulatory processes is vital in shaping EFL learners' early experiences with AI-assisted learning. Practical implications for educators and researchers are discussed, with an emphasis on promoting user-friendly and effective AI environments to support the development of adaptive learning behaviors.

**Keywords:** AI-education, AI-Assisted Learning, Technology Acceptance, Learning Motivation, Metacognitive Strategies, Learning Resilience

## 1. Introduction

The rapid development of artificial intelligence (AI) technologies has led to their increasing integration into language learning environments, particularly in the field of English as a Foreign Language (EFL). AI learning tools are now being employed to support various aspects of language acquisition by enhancing instructional delivery, providing personalized feedback, and facilitating interactive learning experiences [1, 2]. In parallel, theoretical frameworks such as the Technology Acceptance Model (TAM) have been widely used to understand user adoption of technological innovations through constructs like perceived usefulness, perceived ease of use, and perceived self-efficacy [3, 4]. Alongside these, positive psychology constructs—namely optimism, psychological resilience, and growth mindset—have garnered attention for their potential role in fostering student engagement and perseverance in challenging learning contexts [5–7]. This study proposes to conceptualize the aforementioned technology-related factors as "technology acceptance constructs" and the positive psychology factors as "adaptive learning constructs" in order to explore their interrelations in an academic setting.

Current literature has predominantly focused on isolated aspects of AI tool adoption in educational settings, often emphasizing either technology acceptance or positive psychological factors without adequately addressing their potential interplay [8, 9]. Although some studies have investigated the influence of perceived usefulness, ease of use, and self-efficacy on learning outcomes [10, 11], there remains a notable gap in understanding how these factors interact with intrinsic motivational elements and metacognitive strategies during the initial stages of EFL learning [12, 13]. Moreover, existing research has often been limited by sample diversity and methodological constraints, leading to an incomplete picture of the chain mediation processes that may underpin the relationship between technology acceptance and learning resilience [14, 15]. Such shortcomings suggest the necessity for more comprehensive models that integrate motivational and metacognitive variables as mediators in the context of AI-assisted language learning.

In response to these identified gaps, the present study aims to investigate the chain mediation effects of learning motivation and metacognitive strategies on the relationship between technology acceptance constructs and adaptive learning constructs among first-year English majors in China [16, 17]. By employing a rigorous structural equation modeling approach and leveraging data from a diverse sample of EFL beginners, this research seeks to provide nuanced insights into the mechanisms through which AI learning tools influence learners' psychological resilience and overall academic performance [18, 19]. The anticipated outcomes of this study are expected to contribute to a more integrated understanding of technology-enhanced language learning, thereby offering practical implications for educators and policymakers striving to optimize AI-supported instructional practices [20].

## 2. Literature Review

### 2.1 Exploring Technology Acceptance Constructs [21]

This section addresses three central constructs—Perceived Usefulness, Perceived Ease of Use, and Perceived Self-Efficacy—that underpin users' acceptance of AI-based learning tools. These constructs are frequently discussed within the Technology Acceptance Model (TAM) framework and have shown significant influence on learners' intentions to adopt and utilize innovative educational technologies.

#### 2.1.1 Perceived Usefulness [22]

Perceived Usefulness (PU) refers to the extent to which a user believes that a particular technology will enhance task performance. Prior research has demonstrated that when students perceive an AI tool as beneficial for their language acquisition, they display higher engagement and improved learning outcomes. In the context of EFL, PU often translates into facilitating more accurate writing, instantaneous feedback on pronunciation, and more targeted practice opportunities. Such perceptions reinforce learners' belief in AI's capacity to optimize their study processes, thereby promoting overall acceptance.

#### 2.1.2 Perceived Ease of Use [23]

Perceived Ease of Use (PEOU) signifies the degree to which an individual deems a system to be free of effort. Within the EFL domain, learners are more inclined to employ AI tools that present intuitive interfaces, straightforward functionalities, and user-friendly features. Positive evaluations of ease of use not only reduce cognitive load but also encourage continued interaction with the technology. When PEOU is high, learners can allocate more cognitive resources to complex linguistic tasks, potentially fostering deeper language proficiency gains.

#### 2.1.3 Perceived Self-Efficacy [24]

Perceived Self-Efficacy (PSE) concerns learners' judgment of their own capabilities to use a given technology effectively. In an AI-assisted language learning environment, high self-efficacy enables students to navigate unfamiliar functionalities and adapt to new instructional methods with greater confidence. This capacity, in turn, promotes persistent effort and resilience when confronted with initial failures or complex problem-solving tasks. Consequently, PSE has been associated with better performance outcomes, as it mediates the link between learner motivation and technology use.

### 2.2 Motivational and Metacognitive Factors [25]

Learning Motivation and Metacognitive Strategies represent internal processes that significantly shape students' engagement with AI tools. These factors are widely studied in educational psychology due to their essential role in guiding cognitive processes and fostering persistence in language learning.

#### 2.2.1 Learning Motivation [26]

Learning Motivation (LM) encompasses the underlying drives—both intrinsic and extrinsic—that influence learners' goal-setting and sustained effort in acquiring new language skills. Intrinsic motivation emerges from personal interest or enjoyment in the learning task, while extrinsic motivation often stems from external incentives such as grades or social recognition. Empirical findings suggest that learners demonstrating robust motivation display greater willingness to engage deeply with AI-based applications, resulting in heightened language proficiency and satisfaction.

#### 2.2.2 Metacognitive Strategies [27]

Metacognitive Strategies (MS) refer to the deliberate planning, monitoring, and evaluation of one's learning process. Effective metacognitive skills enable students to select appropriate AI tools, set realistic objectives, and adjust their study approaches based on feedback generated by intelligent algorithms. In EFL settings, learners employing higher-order metacognition tend to optimize AI's affordances, such as personalized vocabulary recommendations, thereby achieving improved task efficiency and knowledge retention.

### 2.3 Adaptive Learning Constructs [28, 32]

Optimism, Psychological Resilience, and Growth Mindset are conceptualized here as adaptive learning constructs, reflecting positive psychological attributes that help learners cope with challenges and maintain progress. Research underscores the importance of these factors in enabling learners to flourish within dynamic and often demanding AI-assisted educational contexts.

#### 2.3.1 Optimism [29]

Optimism denotes a general tendency to hold positive expectations for future outcomes. In language learning, optimistic students are more likely to perceive AI technologies as supportive resources, thus investing greater effort in exploring AI-based exercises and feedback mechanisms. Studies imply that optimism fosters a sense of control and encourages learners to persist, even when confronted with complex tasks in EFL environments.

**2.3.2 Psychological Resilience [30]**

Psychological Resilience (PR) encapsulates the capacity to recover from setbacks and adapt effectively to difficulties. In the realm of AI-assisted language learning, resilient students handle technical disruptions, unexpected errors, or less-than-desired initial results by actively seeking alternative solutions or modifying their learning strategies. Over time, such adaptive responses cultivate stronger technology-related coping skills and reinforce learners' confidence in meeting future linguistic and technological challenges.

**2.3.3 Growth Mindset [31]**

A Growth Mindset (GM) frames intelligence and ability as malleable qualities that can be developed through sustained effort and effective strategies. Learners who endorse this belief are more inclined to regard AI feedback—whether corrective or affirmative—as an opportunity for improvement rather than an indicator of fixed competence. As a result, they often demonstrate perseverance, a willingness to experiment with different features, and heightened collaboration with peers to refine their language skills in EFL contexts.

## 3. Research Hypotheses

Figure 1 illustrates a second-order conceptual model, wherein learning motivation (LM) and metacognitive strategies (MS) are posited to exert chain mediation effects on the relationships among perceived usefulness (PU), perceived ease of use (PEOU), perceived self-efficacy (PSE), and learning resilience (operationalized through optimism (OP), psychological resilience (PR), and growth mindset (GM)). Drawing on prior theoretical and empirical findings, the following hypotheses are proposed:

H1: LM and MS jointly mediate the relationship between PU and OP (path ade).
H2: LM and MS jointly mediate the relationship between PU and PR (path adf).
H3: LM and MS jointly mediate the relationship between PU and GM (path adg).
H4: LM and MS jointly mediate the relationship between PEOU and OP (path de).
H5: LM and MS jointly mediate the relationship between PEOU and PR (path bdf).
H6: LM and MS jointly mediate the relationship between PEOU and GM (path bdg).
H7: LM and MS jointly mediate the relationship between PSE and OP (path cde).
H8: LM and MS jointly mediate the relationship between PSE and PR (path cdf).
H9: LM and MS jointly mediate the relationship between PSE and GM (path cdg).

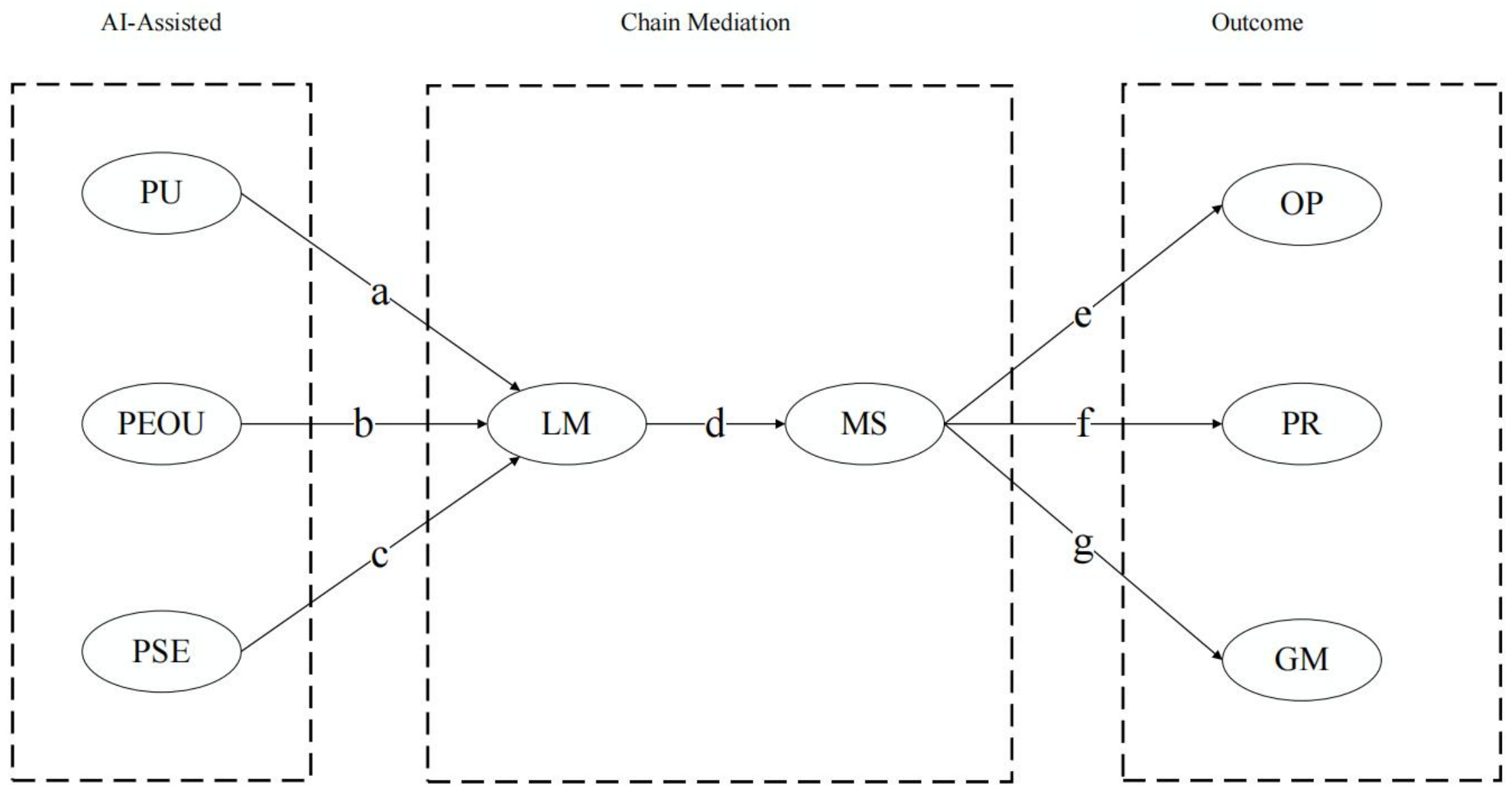

**Figure 1.** Study model diagram (model order: second order)

These hypotheses collectively examine whether the motivational and metacognitive processes link technology acceptance constructs (PU, PEOU, and PSE) to various facets of learning resilience (OP, PR, and GM) among first-year EFL students.

## 4. Methodology

### 4.1 Measurement Instruments

To investigate how AI-assisted language learning influences learning resilience among Chinese English-major undergraduates, and to examine the chain mediation effects of learning motivation and metacognitive strategies, this study employed nine adapted questionnaires. Each questionnaire was selected based on established scales and then modified to fit the context of AI-assisted EFL learning [33–35]. The instruments are presented below in alignment with the constructs of perceived usefulness, perceived ease of use, perceived self-efficacy, learning motivation, metacognitive strategies, optimism, psychological resilience, and growth mindset.

#### 4.1.1 Perceived Usefulness (PU)

The Perceived Usefulness scale was adapted from Siagian et al. [36] to capture students' beliefs regarding whether AI tools could enhance their English learning performance. Items address how effectively AI shortens task completion times, boosts language abilities, and improves overall course engagement. All items employ a 7-point Likert-type format, ranging from 1 (completely disagree) to 7 (completely agree).

#### 4.1.2 Perceived Ease of Use (PEOU)

The Perceived Ease of Use scale was based on Davis's work [37], originally formulated to measure users' perceptions of ease in utilizing information technologies. This study's adapted items focus on whether AI learning tools are straightforward, require minimal effort to master, and reduce cognitive load for EFL learners. Participants rated each statement on a 7-point Likert scale.

#### 4.1.3 Perceived Self-Efficacy (PSE)

Adapted from Tsai et al. [38], this questionnaire measures students' confidence in managing their English learning through AI. It includes items on one's perceived capacity to navigate AI platforms, monitor progress, and adjust study approaches in response to automatic feedback. All items are answered on a 7-point Likert scale, and higher scores denote stronger self-efficacy in AI-mediated EFL settings.

#### 4.1.4 Learning Motivation (LM)

Drawing from Hwang et al. [39], Wang and Chen [40], and Zhu [41], the Learning Motivation scale captures both intrinsic and extrinsic motivational factors in AI-assisted language study. Items reflecting intrinsic motivation (e.g., finding AI-mediated tasks inherently interesting) and extrinsic motivation (e.g., aiming for better grades or future career prospects) were blended to represent a spectrum of underlying motives. Participants responded using a 7-point Likert-type scale.

#### 4.1.5 Metacognitive Strategies (MS)

The Metacognitive Strategies scale, adapted from Wells and Cartwright-Hatton [42], evaluates learners' planning, monitoring, and self-regulation processes in the context of AI-based EFL tasks. The items address cognitive confidence, positive beliefs, and self-awareness while employing AI applications. This scale enables a detailed examination of how students deliberately reflect on and adjust their learning strategies based on real-time AI feedback.

#### 4.1.6 Optimism (OP)

Adapted from Pedrosa et al. [43], the Optimism questionnaire measures students' tendency to hold positive expectations about achieving their English learning goals with AI support. Participants assess the likelihood of overcoming linguistic obstacles and the degree to which they anticipate beneficial learning outcomes. Items use a 7-point Likert-type format, with higher scores indicative of stronger optimism levels.

#### 4.1.7 Psychological Resilience (PR)

The Psychological Resilience scale was adapted from Hu and Gan [44] to explore students' emotional regulation, target focus, and coping behaviors when confronted with setbacks during AI-assisted learning. The questionnaire includes dimensions such as goal clarity, emotional control, and interpersonal support, all scored on a 7-point Likert scale. Higher scores indicate greater resilience in adapting to challenges arising in AI-based EFL tasks.

#### 4.1.8 Growth Mindset (GM)

Finally, the Growth Mindset scale is an adaptation of Sigmundsson and Haga's work [45]. Items measure the extent to which students believe that their English proficiency can be cultivated through sustained effort and the effective use of AI tools. This 7-point Likert-based questionnaire captures learners' perspectives on practice, AI-assisted feedback, and the desire to confront new challenges in pursuit of continuous improvement.

All questionnaires underwent minor linguistic and contextual modifications to reflect the setting of AI-assisted EFL learning. Pilot testing was conducted to ensure clarity and internal consistency. Responses from the pilot participants confirmed that the adapted scales were comprehensible and relevant to the research context.

### 4.2 Investigated Population

A convenience and snowball sampling strategy was employed to recruit newly enrolled English-major undergraduates from 21 provinces in China who were using AI-based digital technologies in their EFL coursework. Following Vanbutsele et al. (2018)[46], a target sample size of five to ten times the total 123 questionnaire items was set. Allowing for non-response and sampling errors [47], 807 questionnaires were distributed. After excluding 75 invalid submissions, 730 valid responses were retained, yielding an effective response rate of 90.46%. Table 1 provides detailed demographic information.

**Table 1.** Demographic information(n=730)

| Demographic Variables | Group | Quantity | Percentage | Demographic Variables | Group | Quantity | Percentage |
|---|---|---|---|---|---|---|---|
| Gender | Male | 372 | 50.96% | Ai tools you have used with high frequency (multiple choice) | Deepseek | 562 | 76.99% |
| | Female | 358 | 49.04% | | ERNIE Bot | 349 | 47.81% |
| Major | Eng. Language and Literature | 163 | 13.2% | | Yuanbao | 221 | 30.27% |
| | Eng. Translation | 106 | 22.1% | | IFlytek Spark | 301 | 41.23% |
| | Eng. Education | 84 | 12.1% | | Chatgpt | 221 | 30.27% |
| | Business Eng. | 71 | 2.4% | | TEMU | 150 | 20.55% |
| | Eng. Linguistics | 28 | 13.0% | | Midjourney | 144 | 19.73% |
| | Interdisciplinary Eng. | 151 | 7.3% | | Tome | 225 | 30.82% |
| | Academic Eng. | 64 | 3.6% | | Runway | 145 | 19.86% |
| | Other Eng. | 63 | 26.3% | | Other | 34 | 4.66% |

# 5. Results

## 5.1 Convergent and Discriminant Validity

Following the recommendation by Marsh et al. (2009)[48], items with factor loadings below 0.50 were removed before conducting confirmatory factor analyses (CFA). As summarized in Appendix 1 and Appendix 2, the standardized factor loadings of the remaining items ranged from 0.715 to 0.987, indicating significant associations between the observed measures and their respective latent constructs. Consistent with Rastegari and Radmehr's (2020)[49] guidelines, composite reliabilities (CR) ranged from 0.828 to 0.951, confirming strong internal consistency among all latent factors. In line with Fornell and Larcker (1981)[50], average variance extracted (AVE) values spanned from 0.547 to 0.651, exceeding the 0.50 threshold and confirming adequate convergent validity. Following this operation, the results of the CFA analyses are summarised in Tables 2 (first order variables) and 3 (second order variables) in Appendices 1 and 2.

Table 4 further demonstrates that the latent variables' means (M) fell between 3.982 and 5.084, suggesting positive evaluations of each construct. Skewness values ranged from 0.029 to 0.378, and kurtosis values from 1.243 to 1.394, meeting standard criteria for normality. Additionally, the square roots of the AVE for each variable exceeded their inter-construct correlations, supporting good discriminant validity across all latent factors. Taken together, these findings indicate that the measurement model achieves satisfactory reliability, convergent validity, and discriminant validity.

**Table 4.** Discriminant Validity Analysis

| | M | SD | Skew | Kurtosis | PU | PEOU | PSE | LM | MS | OP | PR | GM |
|---|---|---|---|---|---|---|---|---|---|---|---|---|
| PU | 4.843 | 1.595 | -0.330 | -1.261 | ***0.784*** | | | | | | | |
| PEOU | 4.304 | 1.179 | -0.125 | -1.258 | .455** | ***0.762*** | | | | | | |
| PSE | 4.503 | 1.626 | -0.211 | -1.303 | .456** | .496** | ***0.786*** | | | | | |
| LM | 3.982 | 1.444 | -0.029 | -1.382 | .486** | .424** | .496** | ***0.782*** | | | | |
| MS | 5.084 | 1.459 | -0.378 | -1.279 | .457** | .474** | .469** | .456** | ***0.764*** | | | |
| OP | 4.597 | 1.622 | -0.238 | -1.243 | .427** | .429** | .425** | .463** | .450** | ***0.789*** | | |
| PR | 4.495 | 1.431 | -0.144 | -1.394 | .476** | .484** | .488** | .501** | .492** | .465** | ***0.766*** | |
| GM | 4.825 | 1.627 | -0.288 | -1.331 | .422** | .456** | .468** | .480** | .488** | .406** | .490** | ***0.791*** |

**: $p < 0.01$, The bold italic represents the square root values of Average Variance Extracted (AVE)

### 5.2 Common Method Bias and Fit Tests

Given the self-reported nature of the data, this study employed anonymous data collection to mitigate potential common method bias. A Harman's single-factor test (Podsakoff et al., 2023) showed that the first factor accounted for 34.576% of the total variance, below the 40% threshold, suggesting that common method bias was not a serious concern. Although a large sample (N = 730) and multiple latent variables can inflate the $\chi^2$ statistic, the overall model fit, as estimated using AMOS 23.0, met recommended benchmarks (Table 5). Notably, $\chi^2$/df = 1.841, GFI = 0.963, AGFI = 0.856, CFI = 0.963, NFI = 0.922, TLI = 0.961, and RMSEA = 0.034, indicating excellent model performance.

**Table 5. Fitted Value**

| $\chi2$ | df | $\chi2$ / df | GFI | AGFI | CFI | NFI | TLI | RMSEA |
|---|---|---|---|---|---|---|---|---|
| 3135.500 | 1703 | 1.841 | 0.963 | 0.856 | 0.963 | 0.922 | 0.961 | 0.034 |

### 5.3 Direct and Chain-Mediated Effects

Table 6 and Figure 2 summarize the results for hypotheses H1 to H9. Each path's significance was assessed via point estimates, standard errors, z-values, bias-corrected 95% confidence intervals, and p-values. A structural equation modeling approach with bootstrapped standard errors was used to derive precise estimates of the direct and mediated effects. As shown in Table 6, the total effect of AI assistance on learning resilience among English majors was 0.588, with a standard error of 0.066, yielding a Z-value of 8.909 (95% CI [0.471, 0.726]), thereby confirming the overall significance of the proposed model.

**Table 6. Mediation Analysis**

| Hypothesis | Mediation path | Point estimate | Product of coefficients | | Bias-Corrected 95% CI | | p | Whether the hypothesis holds |
|---|---|---|---|---|---|---|---|---|
| | | | S.E. | Z | Lower | Upper | | |
| H1 | PU→LM→MS→OP | 0.076 | 0.013 | 5.846 | 0.052 | 0.104 | 0.001 | Yes |
| H2 | PU→LM→MS→PR | 0.062 | 0.010 | 6.200 | 0.042 | 0.083 | 0.001 | Yes |
| H3 | PU→LM→MS→GM | 0.082 | 0.014 | 5.857 | 0.056 | 0.110 | 0.001 | Yes |
| H4 | PEOU→LM→MS→OP | 0.055 | 0.015 | 3.667 | 0.029 | 0.086 | 0.001 | Yes |
| H5 | PEOU→LM→MS→PR | 0.045 | 0.012 | 3.750 | 0.023 | 0.070 | 0.001 | Yes |
| H6 | PEOU→LM→MS→GM | 0.059 | 0.016 | 3.688 | 0.031 | 0.094 | 0.001 | Yes |

| | | | | | | | |
|---|---|---|---|---|---|---|---|
| H7 | PSE→LM→MS→OP | 0.073 | 0.012 | 6.083 | 0.051 | 0.100 | 0.001 | Yes |
| H8 | PSE→LM→MS→PR | 0.059 | 0.010 | 5.900 | 0.041 | 0.080 | 0.001 | Yes |
| H9 | PSE→LM→MS→GM | 0.078 | 0.013 | 6.000 | 0.054 | 0.104 | 0.001 | Yes |
| | total | 0.588 | 0.066 | 8.909 | 0.471 | 0.726 | 0.001 | Yes |

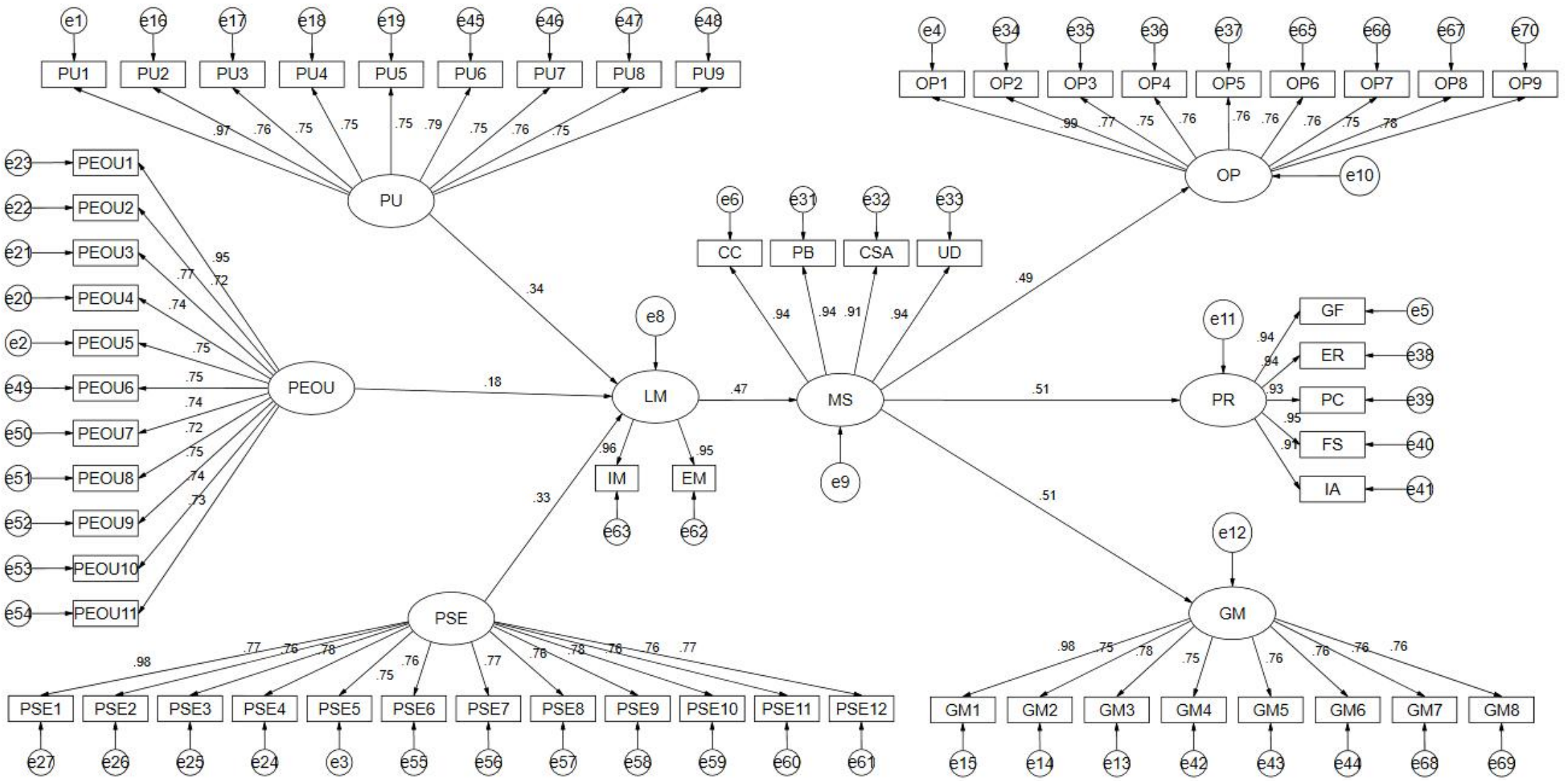

**Figure 2:** Chain mediator model diagram (model order: second order)

## 6. Discussion

The findings of this study underscore the importance of integrating technology acceptance constructs—perceived usefulness (PU), perceived ease of use (PEOU), and perceived self-efficacy (PSE)—with motivational and metacognitive processes to better understand learning resilience in AI-assisted EFL contexts. Specifically, learning motivation (LM) and metacognitive strategies (MS) emerged as significant chain mediators that effectively transmit the positive influence of technology acceptance factors to core resilience constructs: optimism (OP), psychological resilience (PR), and growth mindset (GM). The structural equation modeling results revealed that all mediation paths (H1–H9) were statistically significant, thereby supporting the contention that technology acceptance exerts a meaningful influence on students' capacity to persevere in language learning when facilitated by robust motivational and metacognitive engagement.

These findings align with prior studies suggesting that technology acceptance can positively shape learners' mindset and attitudes, especially when combined with intrinsic and extrinsic motivation [51]. The observed relationships also resonate with research indicating that metacognitive strategies serve as a vital bridge between students' perceived ability to use new technologies and their ultimate resilience outcomes in challenging learning environments [52]. In this study, the total effect of AI assistance on learning resilience ($\beta$ = 0.588) provides empirical evidence that underscores the synergistic role of PU, PEOU, and PSE in motivating deeper engagement. Moreover, the chain mediation mechanism offers a nuanced explanation of how learners' motivational states, paired with adaptive self-regulation techniques, can convert favorable perceptions of AI tools into actionable steps for coping with academic stress and setbacks.

Such a framework highlights the potential of an integrated approach: technology acceptance constructs may create the conditions under which learners feel both competent and inclined to exploit the full range of AI tools, and LM and MS subsequently transform these conditions into resilient behaviors [53]. The consistency of these results with existing literature reinforces the need for future research to broaden their scope across diverse linguistic and cultural settings. Additionally, educators and instructional designers could leverage these insights by focusing on enhancing students' perceived utility of and confidence in AI tools, while simultaneously nurturing intrinsic motivation and metacognitive competence. By doing so, they may foster a cycle of sustained engagement and adaptive learning outcomes, ultimately contributing to the cultivation of well-rounded, resilient EFL learners.

## 7. Conclusion

This study has investigated how AI-assisted learning interventions, mediated by learning motivation (LM) and metacognitive strategies (MS), contribute to the development of optimism (OP), psychological resilience (PR), and growth mindset (GM) among first-year English majors. Drawing on the Technology Acceptance Model (TAM) and positive psychology theories, the research outcomes affirm that perceived usefulness (PU), perceived ease of use (PEOU), and perceived self-efficacy (PSE) collectively serve as crucial catalysts in promoting academic resilience in EFL contexts. The results demonstrate that favorable perceptions of AI tools trigger learning motivation and strategic self-regulation, which, in turn, enhance learners' overall adaptability and perseverance when confronted with linguistic challenges.

In line with theoretical expectations, PU, PEOU, and PSE directly influenced learners' resilience, and these relationships were magnified by the chain mediation of LM and MS. These empirical insights extend existing literature by illustrating how technology acceptance constructs can influence not just performance outcomes, but also core psychological factors conducive to long-term success [54]. Additionally, the observed significance of LM and MS underscores the multi-dimensional nature of the learning process, where both motivational drives and reflective strategies collaborate to shape students' coping mechanisms in AI-supported environments. This comprehensive model indicates that future pedagogical efforts should not only prioritize the technical design and ease of use of AI platforms but also systematically integrate motivational elements and metacognitive training to optimize learning resilience [55].

Taken together, these findings have significant implications for practitioners and policymakers striving to enhance EFL education through AI. Institutions can allocate resources toward developing user-friendly AI systems that reinforce learners' self-efficacy, while also embedding opportunities for goal setting, monitoring, and self-reflection. Such interventions can prime learners to harness the potential of AI-based platforms as catalysts for sustained language development and adaptive learning behaviors. The utility of this integrated approach may further extend beyond EFL settings to other subject domains, where technology acceptance, motivation, and metacognitive regulation similarly interact to influence learners' academic resilience. Future studies may explore cross-cultural comparisons, longitudinal effects, and additional contextual variables such as peer support or instructor feedback, building upon the foundational framework established here.

## Appendix 1

**Table 2.** Confirmatory Factor Analysis Results Table（First order variable）

| Latent Variable | Observed Ariable | Unstd. | SE | Z-Value | P | Std. | SMC | CR | AVE | Cronbach' α |
|---|---|---|---|---|---|---|---|---|---|---|
| PU | PU1 | 1.000 | | | | 0.975 | 0.951 | 0.934 | 0.615 | 0.934 |
| | PU2 | 0.803 | 0.027 | 29.462 | *** | 0.758 | 0.575 | | | |
| | PU3 | 0.753 | 0.026 | 28.453 | *** | 0.746 | 0.557 | | | |
| | PU4 | 0.792 | 0.027 | 29.165 | *** | 0.755 | 0.570 | | | |
| | PU5 | 0.762 | 0.027 | 28.507 | *** | 0.746 | 0.557 | | | |
| | PU6 | 0.841 | 0.026 | 31.98 | *** | 0.787 | 0.619 | | | |
| | PU7 | 0.770 | 0.027 | 28.713 | *** | 0.749 | 0.561 | | | |
| | PU8 | 0.791 | 0.026 | 29.948 | *** | 0.764 | 0.584 | | | |
| | PU9 | 0.751 | 0.026 | 28.743 | *** | 0.749 | 0.561 | | | |
| PEOU | PEOU1 | 1.000 | | | | 0.946 | 0.895 | 0.938 | 0.580 | 0.937 |
| | PEOU2 | 0.791 | 0.031 | 25.726 | *** | 0.723 | 0.523 | | | |
| | PEOU3 | 0.821 | 0.029 | 28.802 | *** | 0.767 | 0.588 | | | |
| | PEOU4 | 0.784 | 0.029 | 26.905 | *** | 0.740 | 0.548 | | | |
| | PEOU5 | 0.830 | 0.03 | 27.346 | *** | 0.747 | 0.558 | | | |
| | PEOU6 | 0.792 | 0.029 | 27.358 | *** | 0.747 | 0.558 | | | |
| | PEOU7 | 0.795 | 0.029 | 27.187 | *** | 0.744 | 0.554 | | | |
| | PEOU8 | 0.801 | 0.031 | 25.59 | *** | 0.721 | 0.520 | | | |
| | PEOU9 | 0.816 | 0.03 | 27.577 | *** | 0.750 | 0.563 | | | |
| | PEOU10 | 0.797 | 0.03 | 26.667 | *** | 0.737 | 0.543 | | | |
| | PEOU11 | 0.776 | 0.03 | 26.013 | *** | 0.727 | 0.529 | | | |
| PSE | PSE1 | 1.000 | | | | 0.98 | 0.960 | 0.951 | 0.618 | 0.950 |
| | PSE2 | 0.808 | 0.026 | 31.022 | *** | 0.77 | 0.593 | | | |
| | PSE3 | 0.767 | 0.025 | 30.285 | *** | 0.762 | 0.581 | | | |
| | PSE4 | 0.814 | 0.026 | 31.908 | *** | 0.779 | 0.607 | | | |
| | PSE5 | 0.789 | 0.027 | 29.618 | *** | 0.754 | 0.569 | | | |
| | PSE6 | 0.778 | 0.026 | 29.924 | *** | 0.758 | 0.575 | | | |
| | PSE7 | 0.808 | 0.026 | 31.381 | *** | 0.774 | 0.599 | | | |
| | PSE8 | 0.798 | 0.026 | 30.326 | *** | 0.762 | 0.581 | | | |
| | PSE9 | 0.812 | 0.026 | 31.602 | *** | 0.776 | 0.602 | | | |
| | PSE10 | 0.804 | 0.027 | 29.888 | *** | 0.757 | 0.573 | | | |
| | PSE11 | 0.811 | 0.027 | 30.347 | *** | 0.762 | 0.581 | | | |
| | PSE12 | 0.806 | 0.026 | 31.185 | *** | 0.772 | 0.596 | | | |
| OP | OP1 | 1.000 | | | | 0.987 | 0.974 | 0.936 | 0.622 | 0.936 |
| | OP2 | 0.821 | 0.026 | 31.162 | *** | 0.769 | 0.591 | | | |
| | OP3 | 0.772 | 0.026 | 29.715 | *** | 0.753 | 0.567 | | | |
| | OP4 | 0.799 | 0.027 | 29.906 | *** | 0.755 | 0.570 | | | |
| | OP5 | 0.812 | 0.027 | 29.951 | *** | 0.756 | 0.572 | | | |
| | OP6 | 0.780 | 0.026 | 30.102 | *** | 0.758 | 0.575 | | | |
| | OP7 | 0.819 | 0.027 | 30.727 | *** | 0.765 | 0.585 | | | |
| | OP8 | 0.776 | 0.027 | 29.185 | *** | 0.747 | 0.558 | | | |
| | OP9 | 0.807 | 0.025 | 31.892 | *** | 0.777 | 0.604 | | | |
| GM | GM1 | 1.000 | | | | 0.978 | 0.956 | 0.930 | 0.626 | 0.929 |
| | GM2 | 0.778 | 0.027 | 28.691 | *** | 0.748 | 0.560 | | | |
| | GM3 | 0.825 | 0.026 | 31.261 | *** | 0.778 | 0.605 | | | |
| | GM4 | 0.782 | 0.027 | 29.039 | *** | 0.752 | 0.566 | | | |
| | GM5 | 0.770 | 0.026 | 29.531 | *** | 0.758 | 0.575 | | | |
| | GM6 | 0.797 | 0.027 | 29.963 | *** | 0.763 | 0.582 | | | |
| | GM7 | 0.778 | 0.026 | 29.935 | *** | 0.763 | 0.582 | | | |
| | GM8 | 0.784 | 0.026 | 29.752 | *** | 0.761 | 0.579 | | | |

***: P<0.001

## Appendix 2

**Table2.** Confirmatory Factor Analysis Results Table（Second order variable）

| LatentVariable | ObservedAriable | Unstd. | SE | Z-Value | P | Std. | SMC | CR | AVE | Cronbach'α |
|---|---|---|---|---|---|---|---|---|---|---|
| LM-IM | IM1 | 1.000 | | | | 0.973 | 0.947 | 0.937 | 0.626 | 0.936 |
| | IM2 | 0.782 | 0.027 | 29.280 | *** | 0.756 | 0.572 | | | |
| | IM3 | 0.803 | 0.027 | 29.931 | *** | 0.764 | 0.584 | | | |
| | IM4 | 0.803 | 0.026 | 31.346 | *** | 0.780 | 0.608 | | | |
| | IM5 | 0.824 | 0.026 | 31.762 | *** | 0.785 | 0.616 | | | |
| | IM6 | 0.786 | 0.027 | 29.573 | *** | 0.760 | 0.578 | | | |
| | IM7 | 0.804 | 0.026 | 30.926 | *** | 0.776 | 0.602 | | | |
| | IM8 | 0.784 | 0.028 | 28.319 | *** | 0.744 | 0.554 | | | |
| | IM9 | 0.785 | 0.027 | 29.220 | *** | 0.756 | 0.572 | | | |
| LM-EM | EM1 | 1.000 | | | | 0.765 | 0.585 | 0.930 | 0.595 | 0.930 |
| | EM2 | 0.993 | 0.045 | 21.968 | *** | 0.775 | 0.601 | | | |
| | EM3 | 0.984 | 0.045 | 21.659 | *** | 0.765 | 0.585 | | | |
| | EM4 | 0.978 | 0.046 | 21.404 | *** | 0.758 | 0.575 | | | |
| | EM5 | 1.007 | 0.046 | 21.916 | *** | 0.773 | 0.598 | | | |
| | EM6 | 1.003 | 0.045 | 22.405 | *** | 0.788 | 0.621 | | | |
| | EM7 | 0.991 | 0.046 | 21.561 | *** | 0.762 | 0.581 | | | |
| | EM8 | 0.968 | 0.045 | 21.723 | *** | 0.767 | 0.588 | | | |
| | EM9 | 1.046 | 0.046 | 22.516 | *** | 0.791 | 0.626 | | | |
| MS-CC | CC1 | 1.000 | | | | 0.959 | 0.920 | 0.896 | 0.636 | 0.893 |
| | CC2 | 0.792 | 0.03 | 26.494 | *** | 0.747 | 0.558 | | | |
| | CC3 | 0.801 | 0.03 | 26.636 | *** | 0.749 | 0.561 | | | |
| | CC4 | 0.773 | 0.029 | 26.598 | *** | 0.749 | 0.561 | | | |
| | CC5 | 0.801 | 0.029 | 27.390 | *** | 0.761 | 0.579 | | | |
| MS-PB | PB1 | 1.000 | | | | 0.756 | 0.572 | 0.893 | 0.582 | 0.893 |
| | PB2 | 1.114 | 0.052 | 21.472 | *** | 0.794 | 0.630 | | | |
| | PB3 | 1.029 | 0.051 | 20.15 | *** | 0.748 | 0.560 | | | |
| | PB4 | 1.067 | 0.051 | 20.938 | *** | 0.775 | 0.601 | | | |
| | PB5 | 1.059 | 0.050 | 21.006 | *** | 0.778 | 0.605 | | | |
| | PB6 | 0.958 | 0.049 | 19.477 | *** | 0.725 | 0.526 | | | |
| MS-CSA | CSA1 | 1.000 | | | | 0.734 | 0.539 | 0.828 | 0.547 | 0.829 |
| | CSA2 | 0.991 | 0.056 | 17.611 | *** | 0.742 | 0.551 | | | |
| | CSA3 | 0.992 | 0.057 | 17.436 | *** | 0.733 | 0.537 | | | |
| | CSA4 | 1.002 | 0.057 | 17.736 | *** | 0.749 | 0.561 | | | |
| MS-UD | UD1 | 1.000 | | | | 0.759 | 0.576 | 0.887 | 0.567 | 0.887 |
| | UD2 | 0.991 | 0.049 | 20.349 | *** | 0.758 | 0.575 | | | |
| | UD3 | 0.988 | 0.049 | 20.242 | *** | 0.754 | 0.569 | | | |
| | UD4 | 0.930 | 0.049 | 19.102 | *** | 0.715 | 0.511 | | | |
| | UD5 | 0.986 | 0.048 | 20.511 | *** | 0.763 | 0.582 | | | |
| | UD6 | 1.032 | 0.05 | 20.680 | *** | 0.769 | 0.591 | | | |
| PR-GF | GF1 | 1.000 | | | | 0.971 | 0.943 | 0.90 | 0.651 | 0.899 |
| | GF2 | 0.783 | 0.028 | 27.551 | *** | 0.752 | 0.566 | 2 | | |
| | GF3 | 0.809 | 0.029 | 27.814 | *** | 0.756 | 0.572 | | | |
| | GF4 | 0.809 | 0.028 | 28.933 | *** | 0.771 | 0.594 | | | |
| | GF5 | 0.772 | 0.027 | 28.357 | *** | 0.763 | 0.582 | | | |
| | ER1 | 1.000 | | | | 0.779 | 0.607 | | | |
| | ER2 | 0.937 | 0.045 | 20.802 | *** | 0.750 | 0.563 | | | |
| PR-ER | ER3 | 1.009 | 0.047 | 21.552 | *** | 0.774 | 0.599 | 0.892 | 0.581 | 0.892 |
| | ER4 | 0.961 | 0.046 | 20.691 | *** | 0.747 | 0.558 | | | |
| | ER5 | 0.979 | 0.046 | 21.317 | *** | 0.766 | 0.587 | | | |
| | ER6 | 0.962 | 0.046 | 20.958 | *** | 0.755 | 0.570 | | | |
| | PC1 | 1.000 | | | | 0.764 | 0.584 | | | |
| PR-PV | PC2 | 0.974 | 0.052 | 18.688 | *** | 0.744 | 0.554 | 0.842 | 0.571 | 0.842 |
| | PC3 | 1.013 | 0.053 | 18.995 | *** | 0.758 | 0.575 | | | |
| | PC4 | 1.022 | 0.054 | 18.977 | *** | 0.757 | 0.573 | | | |
| | FS1 | 1.000 | | | | 0.729 | 0.531 | | | |
| | FS2 | 1.124 | 0.055 | 20.265 | *** | 0.784 | 0.615 | | | |
| PR-FS | FS3 | 1.110 | 0.056 | 19.986 | *** | 0.774 | 0.599 | 0.892 | 0.580 | 0.892 |
| | FS4 | 1.108 | 0.055 | 19.970 | *** | 0.773 | 0.598 | | | |
| | FS5 | 1.084 | 0.055 | 19.766 | *** | 0.765 | 0.585 | | | |
| | FS6 | 1.061 | 0.055 | 19.214 | *** | 0.744 | 0.554 | | | |
| | IA1 | 1.000 | | | | 0.739 | 0.546 | | | |
| PR-IA | IA2 | 1.016 | 0.057 | 17.764 | *** | 0.746 | 0.557 | 0.829 | 0.548 | 0.829 |
| | IA3 | 0.949 | 0.054 | 17.639 | *** | 0.739 | 0.546 | | | |
| | IA4 | 0.989 | 0.056 | 17.584 | *** | 0.736 | 0.542 | | | |

***: P<0.001

## Ethical Approval

This study was approved by the Institutional Review Board at Sichuan Technology and Business University. All research was carried out in accordance with relevant guidelines and regulations (Declaration of Helsinki).

## Informed Consent

Written informed consent was obtained from all participants prior to data collection. Participants were informed that their participation was voluntary, that their responses would remain confidential, and that they could withdraw at any point without penalty. They consented to the publication of anonymized excerpts of their responses in this study.

## Data Availability

Data available on request from the authors.

## Competing Interests

No potential conflicts of interest were declared by the author(s).

## Funding

This research was supported by Kunsan National University's Industry-Academia Cooperation Group (Grant No. 2023H052).